 \documentclass[
twocolumn, amsmath, amssymb, floatfix,
nofootinbib,wrapfloat byrevtex]{revtex4}

 \usepackage{amsmath,mathrsfs,graphicx,epstopdf,placeins,float}
 \usepackage{booktabs}
 \usepackage{multirow}
 \usepackage{pstricks}

 \newcommand{\beq}[1]{\begin{equation}\label{#1}}
 \newcommand{\eeq}{\end{equation}}
 \newcommand{\bea}[1]{\begin{eqnarray}\label{#1}}
 \newcommand{\eea}{\end{eqnarray}}

\begin{document} 

 \title{Holographic complexity growth for a charged AdS-dilaton black holes with fixed and dynamical boundary respectively}
 \author{Ai-chen Li $^{a,b}$ }
 \email{lac@emails.bjut.edu.cn}
 \email{aichenli@visit.uaveiro.eu}
 \affiliation{\it ${}^a$ Departamento de matematica da Universidade de Aveiro and CIDMA,\\ Campus de Santiago, 3810-183 Aveiro, Portugal}
 \affiliation{\it ${}^b$ Theoretical Physics Division, College of Applied Sciences, Beijing University of Technology } 
 
 \begin{abstract}
The holographic complexity conjectures are considered in a Einstein-Maxwell-Dilaton gravity, by using the "Complexity-Volume" proposal. Specifically, we calculate the growth rate of complexity for an eternal charged AdS-dilaton black holes with fixed and dynamical boundaries respectively. The dynamical boundary is achieved by introducing a moving self-graviting brane on which the induced metric has an exact FLRW form. In case of fixed AdS boundary, there exists a bound for evolution of growth rate on late time, while this bound will become larger as the dilaton coupling constant $\alpha$ increases. In large $\alpha$ limit, we analytically prove that this bound is a finite value which is proportional to the black hole mass. In case of dynamical boundary, namely the brane-bulk system, the growth rate decreases monotonously on late time, after reaching a maximum value at a certain time. We find that the evolution of growth rate for brane-bulk system on late time is dominated by the velocity of the moving brane. We guess this result is model-independent.
 \end{abstract}
\maketitle

\section{Introduction}

The Anti-de Sitter/Conformal field theory (AdS/CFT) correspondence \cite{Maldacena:1997re, Witten:1998qj, Gubser:1998bc} provides a profound way for us to associate the physics of gauge field theory with the spacetime geometry in higher dimension. On the one hand, this method allows us to solve a tough problem in strong coupled quantum system by transforming it into a tractable one in classical or semi-classical gravity theory\cite{Hartnoll:2008vx, Gubser:2008yx, Hartnoll:2009sz}. On the other hand, it also pushes us to understand how spacetime geometry emerges from the dual CFT theory. In particular, an intrinsic correlation between the quantum information theory and gravity has been found through the research of holographic entanglement entropy (HEE) \cite{Ryu:2006bv}. In scenario of HEE, the entanglement entropy in boundary field theory is dual to the minimal area of a codimension-2 surface in bulk anchored at the boundaries. Since then, many attempts toward reconstructing the bulk geometry from measures of quantum entanglement have been proposed\cite{Casini:2011kv}.

Recently, another physical quantity called quantum complexity has been  involved into the AdS/CFT dictionary as a complementary concept to HEE. Specifically, in regard to a thermo-field double state (TFD state) which is dual to an eternal asymptotic-AdS black hole \cite{Maldacena:2001kr}, \cite{Hartman:2013qma} shows that the entanglement entropy fails to characterize the growth behavior after reaching the thermal equilibrium for the Einstein-Rosen Bridge (ERB) behind the horizon. Actually, the growth behavior of ERB in black hole interior could be related to the computational complexity of the quantum state on AdS boundary. This insightful view is proposed by the \cite{Stanford:2014jda, Susskind:2014rva} at the first time, namely the so called complexity-volume (CV) conjecture which assumes that the quantum complexity of the CFT on boundary is dual to the maximum volume of the ERB in bulk spacetime. After that, another version called complexity-action (CA) conjecture in context of AdS/CFT is also considered by \cite{Brown:2015bva, Brown:2015lvg}. In the CA conjecture, the quantum complexity on boundary is associated to the gravitational action evaluated on a region of Wheeler-DeWitt (WDW) patch in the bulk spacetime. Basing on these works, the research about quantum complexity from the viewpoint of holography have attracted more and more attentions. Many extensive studies on CV and CA conjectures have been investigated in various aspects, such as the generalization to the charged asymptotic-AdS black holes \cite{Cai:2016xho, Cai:2017sjv} and modified gravity models \cite{Alishahiha:2017hwg, Cano:2018aqi, Jiang:2018pfk, An:2018dbz, Miao:2017quj}, the growth rate \cite{Brown:2017jil, Carmi:2017jqz, Mahapatra:2018gig,Pan:2016ecg}, the divergence structure \cite{Carmi:2016wjl, Reynolds:2016rvl}.

Most works about growth rate of holographic complexity focus on the case of static bulk spacetime and the fixed boundary. Actually, it is also an interesting research direction to explore the time dependence of holographic complexity in a dynamical background spacetime. There have some interesting works on this issue, such as the evolution of holographic complexity in the AdS-Vaidya spacetime \cite{Chapman:2018dem, Chapman:2018lsv, Jiang:2018tlu}. Meanwhile, the \cite{Reynolds:2017lwq} generalizes the complexity conjectures to de-Sitter (dS) space by constructing the dS invariant states in a quantum field theory on boundary and considering the correspondingly holographic calculation in bulk. More interesting, \cite{Caginalp:2019fyt, An:2019opz, Pan:2020kdl} have investigated the growth rate of holographic complexity in a static AdS bulk with a dynamic boundary on which the induced metric has the Friedman-Lemaitre-Robertson-Walker (FLRW) form.

In this paper, our purpose is to consider the holographic complexity conjectures in a Einstein-Maxwell-Dilaton gravity\cite{Gao:2004tv, Gao:2005xv} which is derived from the low-energy limit of string theory\cite{Gibbons:1987ps}. Specifically, we will investigate the growth rate of holographic complexity via the CV proposal for an eternal charged AdS-dilaton black holes with fixed boundaries and  dynamical boundares respectively. The dynamical boundary is achieved by introducing a moving codimension-1 Randall-Sundrum (RS) brane\cite{Randall:1999ee, Randall:1999vf, Chamblin:1999ya, Kraus:1999it}. As we know, for most of asymptotic-AdS black holes, the growth rate of holographic complexity usually increases on early time and then approaches a limit value on late time. And we wonder if the similar physical behavior could be observed in case of the dynamical boundary. Besides, it has been indicated that the dilaton coupling constant play an important role in many phenomenological aspects of dilaton gravity. For example, the different phases structures of black hole thermodynamics is analyzed by \cite{Sheykhi:2009pf,Li:2017kkj} in different dilaton coupling constants; \cite{Xu:2019abl, Li:2019jal} observe the various evolution modes of brane universe as varying the value of dilaton coupling constant. Thus, it is worthwhile to explore the effects of dilaton coupling constant on the growth rate of holographic complexity. In comparison with the work \cite{An:2019opz}, except to generalize their results from AdS-Schwarzschild black hole to charged AdS-dilaton black holes, we also consider the self-gravitating effects of brane by using the method provided in \cite{Li:2019jal, Shiromizu:1999wj, Maeda:2003vq}.

Our work is organized as follows. In Sec.\ref{BHpart}, we briefly review the exact solutions of five-dimensional asymptotically AdS black holes in a type of Einstein-Maxwell-Dilaton gravity. The black hole mass is obtained by using the counterterm method. In Sec.\ref{CVdualBulk}, we investigate the growth rate of holographic complexity for this charged AdS-dilaton black holes by using the CV conjecture. We plot the full time dependence of the growth rate of complexity with different dilaton coupling constant $\alpha$. Besides, the bound of growth rate on late time is calculated analytically in the large $\alpha$ limit. In Sec.\ref{CVFLRWBrane}, we replace the AdS boundary of black holes by a moving self-graviting brane, meanwhile the growth rate of holographic complexity for this brane-bulk system is considered. Finally, Sec.\ref{ConAndDis} will summarize our results and give a discussion.

\section{AdS dilaton black holes and relevant thermodynamical quantities \label{BHpart}}

Let us begin with the action of 5-dimensional Einstein-Maxwell-dilaton gravity (we use the convention $\kappa^2=8\pi G$)
\bea{EMD}
&&\hspace{-5mm}S_{EMD}=\frac{1}{2\kappa_{5}^{2}}\int_{M}d^{5}x\sqrt{-g}[\mathcal{R}-\frac{4}{3}g^{MN}\partial_{M}\phi\partial_{N}\phi
\\
&&\rule{15mm}{0pt}-V(\phi)-e^{-\frac{4}{3}\alpha\phi}F^{2}]
\nonumber
\eea
where $\mathcal{R}$ and $\phi$ are the usual Ricci scalar and dilaton field respectively, the latter has self-interaction $V(\phi)$ and non-minimally couples to the electromagnetic field of kinetic energies $F^2$. The physical constants $\alpha$ measures the strength of this coupling. Equation of motions following from $\eqref{EMD}$ have the form
\bea{feq}
\nonumber
&&\hspace{-5mm}\mathcal{R}_{MN}=\frac{1}{3}[4\partial_{M}\phi\partial_{N}\phi+g_{MN}V(\phi)]+2e^{-\frac{4\alpha\phi}{3}}[F^L _M F_{LN}                               
\\
&&\hspace{-5mm} \quad \quad \quad -\frac{1}{6}g_{MN}F^{2}] 
\eea
\bea{seq}
&&\hspace{-11mm} \nabla^{2}\phi=\frac{\partial_{M}(\sqrt{-g}g^{MN}\partial_{N}\phi)}{\sqrt{-g}}=\frac{3}{8}\frac{\partial V}{\partial\phi}-\frac{\alpha}{2}e^{-\frac{4\alpha\phi}{3}}F^{2}  
\eea
\bea{Maxwell}
&&\hspace{-11mm} \nabla_{N}(e^{-\frac{4\alpha\phi}{3}}F^{NM})=\partial_{N}(\sqrt{-g}e^{-\frac{4\alpha\phi}{3}}F^{NM})=0
\eea
We consider a static black hole solution with metric ansatz
\bea{Metric}
\nonumber
&& ds^2 =g_{AB}dx^Adx^B \\
\label{Ge5dBHAnsa}
&& \quad ~=-A(r)dt^2+B(r)dr^2+R(r)^2d\Omega ^2 _{k,3}
\eea
where $d\Omega ^2 _{k,3}$ is the line element of 3-dimensional hyper surface of constant curvature $6k$ with $k=\pm1, 0$ corresponding to spheric, hyperbolic and plane topology respectively. For simplicity, we will consider only the spherical case in this paper, namely $k=1$ and $d\Omega ^2 _3 =d\theta ^2 _1+\sin^2\theta_1 d\theta^2 _2+\sin^2\theta_1 \sin^2 \theta_2 d\varphi ^2$. We assume only static electric fields exists in this system, and then the only nonzero components of electromagnetic field strength is $F_{tr}=-F_{rt}$. According to the maxwell equation \eqref{Maxwell}, $F_{tr}$ is solved as
\bea{solEM}
F_{tr} =\sqrt{A(r)B(r)}\frac{qe^{4\alpha\phi/3}}{R^{3}(r)}
\eea
Explicitly, substitute \eqref{Metric},\eqref{Maxwell} into \eqref{feq} and \eqref{seq}, it can be obtained that
\bea{}
\label{Eintt}
&&\hspace{-14mm}\frac{A''}{2B}+\frac{3A'R'}{2BR}-\frac{A'B'}{4B^{2}}-\frac{(A')^{2}}{4AB}=\frac{4e^{-4\alpha\phi/3}(F_{tr})^{2}}{3B}-\frac{1}{3}AV
\\
\label{Einrr}
&&\hspace{-14mm}\frac{A''}{2A}+\frac{3R''}{R}-\frac{3B'R'}{2BR}-\frac{A'B'}{4AB}-\frac{(A')^{2}}{4A^{2}}=-\frac{1}{3}BV-\frac{4}{3}(\phi')^{2}\\
\nonumber
&&\hspace{-4mm}\quad\quad\quad\quad\quad\quad\quad\quad\quad\quad\quad\quad\quad\quad +\frac{4(F_{tr})^{2}e^{-4\alpha\phi/3}}{3A}
\\
\label{Einxx}
&&\hspace{-14mm}\frac{RR''}{B}+\frac{2(R')^{2}}{B}-\frac{B'R'R}{2B^{2}}+\frac{A'R'R}{2AB}=-\frac{1}{3}R^{2}V
\\
\nonumber 
&&\hspace{-4mm}\quad\quad\quad\quad\quad\quad\quad\quad\quad\quad\quad\quad\quad-\frac{2R^{2}(F_{tr})^{2}e^{-4\alpha\phi/3}}{3AB}
\\
\label{eqphi}
&&\hspace{-14mm}\frac{\phi''}{B}+\frac{3R'\phi'}{BR}-\frac{B'\phi'}{2B^{2}}+\frac{A'\phi'}{2AB}=\frac{3}{8}V'+\frac{\alpha(F_{tr})^{2}e^{-4\alpha\phi/3}}{AB}
\eea
By adjusting the form of $V(\phi)$ appropriately, reference \cite{Gao:2004tv, Gao:2005xv} obtains asymptotically AdS black hole solutions of the system analytically,
\bea{}
\label{solDilaPon}
&&\hspace{-3mm}V(\phi)=\frac{\Lambda}{2(2+\alpha^{2})^{2}}(4\alpha^{2}(\alpha^{2}-1)\cdot e^{-\frac{8\phi}{3\alpha}}\\
\nonumber
&& \hspace{-3mm}\quad \quad ~ +4(4-\alpha^{2})\cdot e^{\frac{4\alpha\phi}{3}}+24\alpha^{2}\cdot e^{-\frac{2(2-\alpha^{2})\phi}{3\alpha}})
\\
\label{solA}
&&\hspace{-3mm}A(r)=(k-\frac{c^{2}}{r^{2}})(1-\frac{b^{2}}{r^{2}})^{\frac{2-\alpha^{2}}{2+\alpha^{2}}}-\frac{\Lambda r^{2}}{6}(1-\frac{b^{2}}{r^{2}})^{\frac{\alpha^{2}}{2+\alpha^{2}}}
\\
\label{solB}
&&\hspace{-3mm}B(r)=(1-\frac{b^{2}}{r^{2}})^{-\frac{\alpha^{2}}{2+\alpha^{2}}}/A(r)
\\
\label{solR}
&&\hspace{-3mm}R(r)=(1-\frac{b^{2}}{r^{2}})^{\frac{\alpha^{2}}{2(2+\alpha^{2})}}r
\\
\label{solPhi}
&& \hspace{-3mm}\phi(r)=\frac{3\alpha}{2(2+\alpha^{2})}\ln(1-\frac{b^{2}}{r^{2}})
\eea
Besides, the following parameter relation is also implied by Einstein field equations,
\bea{}
\label{ElectriWithbc}
q^{2}=\frac{6}{(2+\alpha^{2})}b^{2}c^{2}
\eea
Here $b$ and $c$ are integration constants with dimension of length, as we show later, they are also related with the mass $M$. Reference \cite{Sheykhi:2009pf} shows that for this black hole solution, the Kretschmann scalar $R^{\mu \nu \alpha \beta} R_{\mu \nu \alpha \beta}$ and the Ricci scalar R both diverge at $r =b$, thus $r =b$ is the location of curvature singularity. Furthermore, it is important to recognize that this charged AdS-dilaton only exists one horizon whatever the value of dilaton coupling constant $\alpha$.

When we use the Euclidean path integral approach to calculate the thermodynamical quantity and black mass, there exists a unavoidable divergence on boundary of spacetime. In asymptotically AdS spacetime, a suitable surface counterterm is found with feature of coordinate frame independence, after the renormalization procedure, we could obtain a finite Euclidean action and a well-defined boundary stress-energy tensor. According to the methods in \cite{Balasubramanian:1999re, Cai:1999xg}, we choose the surface counterterm as the following ansatz
\bea{}
\nonumber
&&\hspace{-2mm} S_{ct}	=	-\frac{1}{\kappa_{5}^{2}}\int_{\partial M}d^{4}x\sqrt{-\gamma}\bigg\{\frac{c_{0}}{l_{eff}(\phi)}(1+\frac{c_{\phi}}{c_{0}}\phi^{2})\\
\label{CounterAction}
&&\hspace{-2mm}\quad \quad \quad +c_{1}l_{eff}\mathcal{R}+c_2l_{eff}^{3}\big(\mathcal{R}^{2}+c_\beta \mathcal{R}^{ab}\mathcal{R}_{ab}\big)\bigg\} \\
\nonumber
\\
\nonumber
&& where\quad  \frac{1}{l_{eff}(\phi)}=\sqrt{-\frac{V(\phi)}{3(3+1)}} 
\eea
the expression of $V(\phi)$ has shown in eqs.$\eqref{solDilaPon}$. Actually, $\mathcal{R}^{2}$ and $\mathcal{R}^{ab}\mathcal{R}_{ab}$ are smae in magnitude, we will make $c_\beta=0$ for simplifying the caculations. With inclusion of this counterterm, the quasilocal stress-energy tensor at the boundary $r=const$ with induced metric $\gamma_{ab}$ could be derived as
\bea{}
\nonumber
&&T_{ab}	=	\frac{1}{\kappa_{5}^{2}}\bigg\{ K_{ab}-K\gamma_{ab}+\frac{c_{0}}{l_{eff}}\big(1+c_{\phi}\phi^{2}\big)\gamma_{ab}\\
\nonumber
&&\quad \quad -2c_{1}l_{eff}\big(\mathcal{R}_{ab}-\frac{1}{2}\mathcal{R}\gamma_{ab}\big)+c_{2}l_{eff}^{3}\big(\gamma_{ab}\mathcal{R}^{2}\\
\label{GeneTab}
&&\quad \quad-4\mathcal{R}\mathcal{R}_{ab}+4\nabla_{a}\nabla_{b}\mathcal{R}-4\gamma_{ab}\nabla_{m}\nabla^{m}\mathcal{R}\big)\bigg\}		
\eea
where the $\gamma_{ab}$ is the induced metric on the boundary $r=const$, which is defined as
\bea{}
\gamma_{ab}dx^adx^b=\lim_{r\to con} ds^2_5=-A(r)dt^2 +\big(R(r) \big) ^2d\Omega^2_3
\eea
The $K_{ab}$ is the extrinsic curvature on the boundary. Expand the $\eqref{GeneTab}$ explicitly, we give
\bea{}
\nonumber
&&T_{tt}=-3\frac{AR^{\prime}}{\sqrt{B}R}-\frac{c_{0}}{l_{eff}}A\big(1+c_{\phi}\phi^{2}\big)\\
\label{WithCouTtt}
&&\quad \quad -6\frac{c_{1}l_{eff}}{R^{2}}A-36\frac{c_{2}l_{eff}^{3}}{R^{4}}A\\
\nonumber
&&T_{ij}	=	\frac{h_{ij}}{R^{2}}\bigg(\frac{R^{2}A^{\prime}}{2A\sqrt{B}}+2\frac{RR^{\prime}}{\sqrt{B}}+c_{0}(1+\frac{c_{\alpha}}{c_{0}}\phi^{2})\frac{R^{2}}{l_{eff}}\\
\label{WithCouTij}
&& \quad \quad +2c_{1}l_{eff}-12c_{2}\frac{l_{eff}^{3}}{R^{2}}\bigg)
\eea
The mass of black hole is a conserved charge associated with a timelike killing vector, from the reference \cite{Balasubramanian:1999re}, it could be defined as
\bea{}
\label{DefiOfBHM}
M=\int _{r\to \infty} dx^3  \big(R(r) \big) ^3 \big( A(r) \big) ^{-\frac{1}{2}} T_{tt}
\eea 
For getting a finite mass of black hole, we need to find a well-defined quasilocal stress-energy $T_{tt}$. Thus, we choose the undetermined coefficients in \eqref{CounterAction} as $c_0=-3, c_\phi=\frac{2}{9}, c_1=-\frac{1}{4}, c_2=\frac{1}{96}$ \cite{Li:2019jal}. After substitute these coefficients and \eqref{solDilaPon}-\eqref{solPhi} into $\eqref{DefiOfBHM}$, the expression of black hole mass is obtained as
\bea{}
\label{FiExOf}
M=\frac{3\Omega_{3}}{2\kappa_{5}^{2}}\bigg(c^{2}+\big(\frac{2-\alpha^{2}}{2+\alpha^{2}}\big)b^{2}\bigg)
\eea

\section{The Growth Rate of Complexity for charged AdS-dilaton black hole via the C-V conjecture \label{CVdualBulk}}

\begin{figure}[ht]
	\begin{center}
		\includegraphics[scale=0.75]{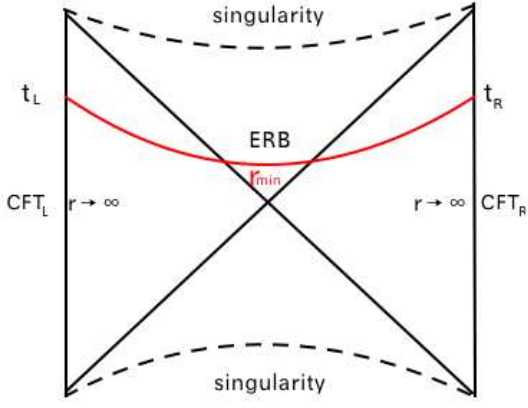}
		\caption{(color online). Penrose diagram for an eternal asymptotic-AdS black hole with one horizon. The CV conjecture implies that the computational complexity of the dual quantum state on spacetime boundary is associated to the maximal size of an Einstein-Rosen bridge (ERB). }
		\label{AdSDilaPenrose}
	\end{center}
\end{figure}

In this section, let us calculate the growth rate of complexity for the charged AdS-dilaton black hole solutions \eqref{} via the complexity=volume (C-V) conjecture, following the standard procedure given by \cite{Stanford:2014jda, Carmi:2017jqz}. The Penrose diagram of an eternal charged AdS-dilaton black hole is shown in Figure \ref{AdSDilaPenrose}, which is similar to the case of an eternal Schwarzschild-AdS black hole. There exists two copies of CFT, denoted by $CFT_L$ and $CFT_R$, living on left and right timelike boundaries respectively. The red curve in Fig.\ref{AdSDilaPenrose} represents the maximal wormhole (codimension-one bulk surface) connecting the desired time slices at $t_L$ and $t_R$ on the two asymptotic boundaries. According to the C-V conjecture, the complexity of the thermofield double state associated to $CFT_L$ and $CFT_R$ on boundary spacetime is dual to the maximal volume of the wormhole, 
\begin{align}
\label{ComplexInCV}
\mathcal{C}_{V}=\frac{8\pi}{\kappa^2 _5 L} \mathcal{V}_{max}
\end{align}
We use the $r_{min}$ to denote the minimum distance which the Einstein-Rosen bridge can reach, while $r_{min}$ is inside the future horizon $r_h$. 
From CFT's side, the thermofield double state is invariant under an evolution with the Hamiltonian $H=H_L-H_R$. Correspondingly, the time dependence of the complexity only depend on the total time $\eta=t_L+t_R$ and not on each of the boundary times respectively. Without loss of generality, we could always calculate the growth rate of holographic complexity for a symmetric configuration, i.e. $t_L=t_R=\frac{\eta}{2}$. Besides, due to the time reversal symmetry $\eta\to-\eta$ of the bulk geometry, we just need to consider the behavior of the complexity in range $r\geq 0$, namely the upper half of the Penrose diagram.

\begin{figure}[ht]
	\begin{center}
		\includegraphics[scale=0.75]{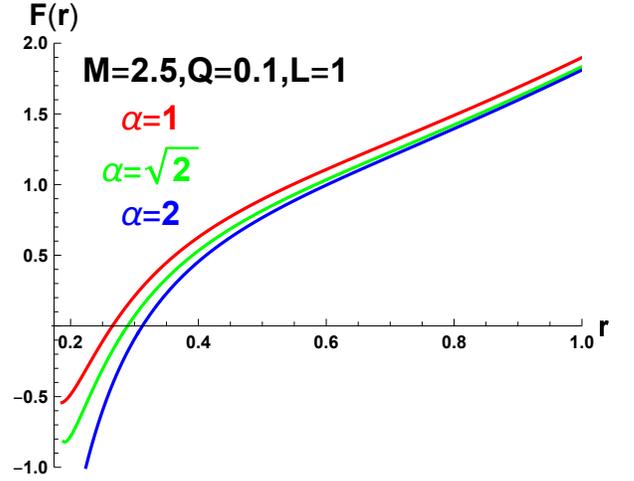}
		\caption{(color online). The variation of $F(r)$ with respect to the $r$ in some representative parameters.}
		\label{PloteffF}
	\end{center}
\end{figure}

Transform the line element $\eqref{}$ into the Eddington-Finkelstein coordinates, we give
\begin{align}
&ds^{2}=-A(r)dv^{2}+\frac{2A(r)}{F(r)}dvdr+R(r)^{2}d\Omega_{k,3}^{2}
\end{align}
where
\begin{align}
\label{DefiInfaMode}
&v=t+r^\star(r) ~,~dr^{\star}=\frac{dr}{F(r)}\\
&F(r)=\begin{cases}
\begin{array}{c}
\sqrt{\frac{A(r)}{B(r)}}~,~r\geq r_{h}\\
-\sqrt{\frac{A(r)}{B(r)}}~,~r \leq r_{h}
\end{array}\end{cases}		
\end{align}
As shown by Fig.\ref{PloteffF}, the $F(r)$ is continuous at horizon $r=r_h$.
Assuming the geometry of wormhole, i.e. the codimension-one bulk surface, has the same maximal symmetry as the horizon geometry of black holes. Thus we just need to embed the surface into the bulk spacetime by $v(\lambda)$ and $r(\lambda)$, in which the parameter $\lambda$ is the radial coordinate intrinsic to the surface. And then the volume of wormhole is expressed as
\begin{align}
\label{VolWorm}
\mathcal{V}=2\Omega_{k,3}\int^{\lambda_{max}} _{\lambda_{min}} d\lambda~ R(r)^{3}\sqrt{-A(r)\dot{v}^{2}+\frac{2A(r)}{F(r)}\dot{v}\dot{r}}
\end{align}
where the dots denote the derivative with respect to $\lambda$. The overall factor $2$ is owing to the fact that the surface is composed of two equivalent parts, while the integral only cover the left or right part. We could obtain the maximal volume by extremizing the $\eqref{VolWorm}$. For the convenience of calculations in later, we rewrite the $\eqref{VolWorm}$ as
\begin{align}
\label{VolLag}
&\mathcal{V}=2\Omega_{k,3}\int d\lambda\mathcal{L}(\dot{v},\dot{r},r)
\end{align}
Since the $\mathcal{L}(\dot{v},\dot{r},r)$ does not depend explicitly on $v$, a conserved quantity $E$ is implied when varying $\eqref{VolLag}$ with respect to $v(\lambda)$, namely
\begin{align}
\label{EOMsv}
&\frac{\partial \mathcal{L}}{\partial \dot{v}}=\sqrt{\frac{A}{F}}\frac{R^{3}(\dot{r}-F\dot{v})}{\sqrt{2\dot{r}\dot{v}-F\dot{v}^{2}}}=-E
\end{align}
it is easy to observe that the $E$ has the dimension of energy. Instead of deriving the other equations of motion about $r(\lambda)$ from $\eqref{VolWorm}$, we will use the fact that the expression in $\eqref{VolWorm}$ is reparametrization invariant. In this way, the $\lambda$ is freely chosen to keep the radial volume element fixed, without loss of generality,
\begin{align}
\label{GaugeFixReInva}
R^{3}\sqrt{\frac{A}{F}}\sqrt{2\dot{v}\dot{r}-F\dot{v}^{2}}=1
\end{align}
We could further simplify $\eqref{EOMsv}$ and $\eqref{GaugeFixReInva}$ as,
\begin{align}
\label{repdotr}
&\dot{r}=\frac{F}{AR^{3}}\sqrt{(A+\frac{E^{2}}{R^{6}})}\\
\label{repdotv}
&\dot{v}=\frac{1}{AR^{3}}\sqrt{(A+\frac{E^{2}}{R^{6}})}+\frac{E}{AR^{6}}
\end{align}
Thus, according to $\eqref{VolWorm}, \eqref{GaugeFixReInva}$ and $\eqref{repdotr}$, the maximal volume of wormhole is expressed as
\begin{align}
\nonumber
&\mathcal{V}_{max}=2\Omega_{k,3}\int_{r_{min}}^{r_{max}}\frac{1}{\dot{r}}dr\\
\label{MaxVolWorm}
&\quad\quad~=2\Omega_{k,3}\int_{r_{min}}^{r_{max}}\frac{AR^{6}}{F\sqrt{(AR^{6}+E^{2})}}dr
\end{align}
As shown by Fig.\ref{AdSDilaPenrose}, since we consider a symmetric configuration, the minimal radius $r_{min}$ should be a turning point of the surface, i.e. $\dot{r}\vert_{r_{min}}=0$. Moreover, the $r_{min}$ is associated with the parameter $E$ through $\eqref{repdotr}$,
\begin{align}
\label{AssoEandrmin}
&-A(r_{min})R(r_{min})^{6}=E^{2}
\end{align}
Note that the equation $\eqref{AssoEandrmin}$ also suggests $r_{min}<r_h$, while we have $ \dot{v}\vert_{r_{min}}>0,~0>E$ from $\eqref{EOMsv}$. At the same time, it is easy to see that the $r_{max}=\infty$, which is the location of spacetime boundary.

By combining the $\eqref{repdotr}$ with $\eqref{repdotv}$, we give
\begin{align}
\nonumber
&\hspace{-1mm}v_{\infty}-v_{min}=\int_{r_{min}}^{\infty}\frac{dv}{d\lambda}\frac{d\lambda}{dr}dr=\int_{r_{min}}^{\infty}\frac{\dot{v}}{\dot{r}}dr\\
\label{ReWritev}
&\quad\quad\quad\quad~=\int_{r_{min}}^{\infty}\big\{\frac{1}{F}+\frac{E}{F\sqrt{(AR^{6}+E^{2})}}\big\} dr
\end{align}
From the definition $\eqref{DefiInfaMode}$, we also have
\begin{align}
\label{ReWritevAnoWay}
&v_{\infty}-v_{min}=t_{\infty}+r^{\star}(\infty)-t_{min}-r^{\star}(r_{min})
\end{align}
actually the time at the turning point can be set to zero, i.e. $t_{min}=0$, due to the symmetry of configurations. Through $\eqref{ReWritev}$ and $\eqref{ReWritevAnoWay}$, the time on spacetime boundary is expressed as
\begin{align}
\label{TimeOnBoun}
t_{\infty}=\int_{r_{min}}^{\infty}\big\{\frac{E}{F\sqrt{(AR^{6}+E^{2})}}\big\} dr
\end{align}
Here the $t_\infty$ only represents the time $t_L$ on the left spacetime boundary or the time $t_R$ on the right spacetime boundary, as shown by Fig.\ref{AdSDilaPenrose}. However, the time dependence of the complexity are only relevant with the total time $\eta=t_L+t_R=2t_\infty$.

\begin{figure}[ht]
	\begin{center}
		\includegraphics[scale=0.75]{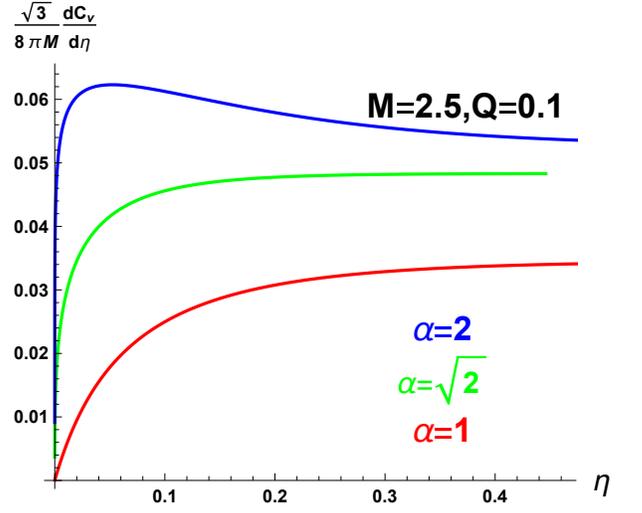}
		\caption{(color online). The full time dependence of the holographic complexity growth for an eternal charged AdS-dilaton black holes in different $\alpha$. The growth rate is converted to a dimensionless quantity by multiplying it by $\frac{\sqrt{3}}{8\pi M}$.}
		\label{PicGrowCom}
	\end{center}
\end{figure}

Associating $\eqref{ComplexInCV},~\eqref{MaxVolWorm}$ to $\eqref{TimeOnBoun}$, the growth rate of complexity is given as
\begin{align}
\nonumber
&\frac{d\mathcal{C}_{V}}{d\eta}=\frac{8\pi}{\kappa^2 _5 L} \frac{d\mathcal{V}_{max}}{d\eta}=\frac{4\pi}{\kappa^2 _5 L}(\frac{d\mathcal{V}_{max}}{dr_{min}})\bigg/(\frac{dt_{\infty}}{dr_{min}})\\
\nonumber
&\quad\quad=\frac{4\pi}{\kappa^2 _5 L}\big(\frac{-2\Omega_{k,3}AR^{6}}{F\sqrt{AR^{6}+E^{2}}}\big)_{r_{min}}\bigg/\big(\frac{-E}{F\sqrt{AR^{6}+E^{2}}}\big)_{r_{min}}\\
\label{ComplexGrowth}
&\quad\quad=\frac{8\pi\Omega_{k,3}\sqrt{-A(r_{min})}R(r_{min})^{3}}{\kappa_{5}^{2}L}
\end{align}
The time dependence of the complexity growth rate for AdS-dilaton black hole solutions $\eqref{solA}-\eqref{solR}$ in case of $k=1$ is shown by Figure \ref{PicGrowCom}. It is easy to observe that the growth rate approaches a constant as the time $\eta$ increases, while this constant is viewed as the bound of complexity growth rate at late times. Besides, we could also find that the larger the dilaton coupling $\alpha$ is, the higher the bound would be. 
Next, let us prove that the bound is a finite value when $\alpha \to \infty$. In the large $\alpha$ limit, the $\eqref{solA}$ and $\eqref{solR}$ can be approximated as,
\begin{align}
\label{solALargAlph}
&A(r)\big\vert_{\alpha \to \infty}=(1-\frac{c^{2}}{r^{2}})(1-\frac{b^{2}}{r^{2}})^{-1}+\frac{r^{2}}{L^{2}}(1-\frac{b^{2}}{r^{2}})\\
\label{solRLargAlph}
&R(r)\big\vert_{\alpha \to \infty}=(1-\frac{b^{2}}{r^{2}})^{\frac{1}{2}}r
\end{align}
Meanwhile, the black hole mass become
\begin{align}
\label{BHMassLarAlph}
&M\big\vert_{\alpha \to \infty}=\frac{3\Omega_{3}}{2\kappa_{5}^{2}}(c^{2}-b^{2})
\end{align} 
From $\eqref{ComplexGrowth}$, it is easy to see that the bound of $\frac{d\mathcal{C}_{V}}{d\eta}$ could be found by searching the maximum value of $\sqrt{-A(r_{min})}R(r_{min})^{3}$. We use $\tilde{r}_{min}$ to denote the extremum point of $\sqrt{-A(r_{min})}R(r_{min})^{3}$, namely
\begin{align}
\label{SearExtrePoint}
&\frac{d}{dr_{min}}\big(\sqrt{-A(r_{min})}R(r_{min})^{3}\big)\bigg\vert_{\tilde{r}_{min}}=0
\end{align}
Plug $\eqref{solALargAlph}$, $\eqref{solRLargAlph}$ into $\eqref{SearExtrePoint}$, we obtain
\begin{align}
&\tilde{r}_{min}^{\pm}=\frac{\sqrt{8b^{2}-3L^{2}\pm L\sqrt{32c^{2}-32b^{2}+9L^{2}}}}{2\sqrt{2}}
\end{align}
We will throw away $\tilde{r}_{min}^{-}$ because of $\tilde{r}_{min}^{-}<b$ ($r=b$ is the singularity of black hole). Furthermore, one can easily check that the $\tilde{r}_{min}^+$ is the maximum point. After substituting $\tilde{r}^+ _{min}$ into $\eqref{ComplexGrowth}$ and using some tricks of inequality, we give
\begin{align}
\label{FinaBound}
\frac{d\mathcal{C}_{V}}{d\eta}\bigg\vert_{\eta \to \infty}<\frac{4\sqrt{3}\pi\Omega_{k,3}(c^{2}-b^{2})}{\kappa_{5}^{2}}=\frac{8\pi M}{\sqrt{3}}
\end{align}
where we have used the $\eqref{BHMassLarAlph}$. In brief, $\eqref{FinaBound}$ means that the complexity growth rate, especially in the late time limit, can not exceed the bound $\frac{8\pi M}{\sqrt{3}}$, whatever how large the dilaton coupling $\alpha$ is.

\section{The Growth Rate of Complexity for brane universe moving in charged AdS-dilaton black holes \label{CVFLRWBrane}}

\subsection{Motion of brane in charged AdS-dilaton black holes}

In this part, we briefly review the movement of a self-graviting 3-brane in charged AdS-dilaton black holes, one can refer \cite{Li:2019jal} to get more detail. We assume $\mathcal{M}$ is a five-dimensional manifold containing a brane $\Sigma$ with two sides, which splits $\mathcal{M}$ into two parts with a $Z_2$ symmetry. Here, $X^A=\{t,r,\vec{x}_3\}$ and $x^\mu=\{\tau,\vec{x}_3\}$ are used to denote bulk coordinates and internal coordinates of brane world sheet respectively. We consider the scenario that the brane moves translationally along the $r$-direction, thus the brane is embedded into the bulk spacetime by $r(\tau)$ and $t(\tau)$, in which $\tau$ is the proper time on brane. For the convenience of distinguishing radial coordinates $r$ and the locaition of brane $r(\tau)$, we replace $r(\tau)$ by symbol $a(\tau)$. The velocity vector of the brane could be written as $u^M=(\dot{t}(\tau), \dot{a}(\tau),0,0,0)$. According to the normalization condition $u^M u_M=-1$, we have
\begin{align}
\label{ttauRela}
&\frac{dt}{d\tau}=\sqrt{(1+B\dot{a}^2)/A}
\end{align}
Let $n_M$ be the unit normal point into $\mathcal{M}$. Through orthogonal condition $u^M n_M=0,~n^M n_M=1$, the $n_M$ could be expressed as
\begin{align}
 n_M=(\sqrt{AB}\dot{a},-\sqrt{B(1+B\dot{a}^2)},0,0,0)
 \end{align}
The induced metric on brane is
\begin{align}
\label{InduMetOnBrane}
ds^2=-d\tau^2+R \big( a(\tau) \big) ^2 d\Omega ^2 _{k,3}\\
\nonumber
h_{\mu\nu}=e^M _\mu e^N _\nu (g_{MN}-n_M n_N)
\end{align}
in which the vierbein $e^M _\mu$ is defined as $e^M _\mu=\frac{\partial X^A}{\partial x^\mu}$. The bulk-brane system is described by the following action
\begin{align}
\label{TotalBulkBrane}
&\quad \quad S=S_{EMD} +S_{brane}\\
&S_{brane}=\int d^4 x\sqrt{-h} \{ \frac{\mathcal{K}}{\kappa^2 _5}+\lambda(\phi)+\mathcal{L}_{matter} \}
\end{align}
where $\mathcal{L}_{EMD}$ is the lagrangian of Einstein-Maxwell-Dilaton gravity. Meanwhile, $\lambda$ represents the effective brane tension, which is an undetermined function of $\phi$. And $\mathcal{K}$ is the trace of the extrinsic curvature, namely $\mathcal{K}=h^{\mu \nu} \mathcal{K}_{\mu\nu}=h^{\mu \nu} e^M _\mu e^N_\nu (\nabla_M n_N+\nabla_N n_M)$. Varying $\eqref{TotalBulkBrane}$ with respect to $g^{\mu\nu}$, except to derive a standard Einstein equation in bulk spacetime, we also obtain an Israel junction condition on the brane
\begin{align}
\label{geneIsrael}
&\mathcal{K}_{\mu\nu}-\mathcal{K}h_{\mu\nu}=\frac{\kappa^2_5}{2}S_{\mu\nu}
\end{align}
in which we have assumed a $Z_2$ symmetry on two sides of the brane, namely $(\mathcal{K_{\mu\nu}})\big\vert_{\Sigma_-}=-(\mathcal{K_{\mu\nu}})\big\vert_{\Sigma_+}$. In a similar way, the boundary condition of dilaton field on the brane is obtained as,
\begin{align}
\label{AbstDilaBouCon}
\frac{4}{3\kappa^2_5} n^M \partial _M \phi =\frac{\partial \lambda}{\partial \phi}
\end{align}
Expanding $\eqref{AbstDilaBouCon}$ yields,
\begin{align}
\label{DetaJunScalar}
&-\frac{\sqrt{(1+B\dot{a}^{2})}}{\sqrt{B}}\partial_{r}\phi=\frac{3\kappa_{5}^{2}}{4}\frac{\partial\lambda}{\partial\phi}
\end{align}
Meanwhile, since the boundary condition of vector field on the brane is trivial, we will not consider it.

Moreover, the energy-momentum tensor of matters on the brane is assumed to have the form of simple ideal fluid,
\begin{align}
\label{Exoidealfluid}
\mathcal{S}^\mu _\nu=Diag\{-\rho,P,P,P \}
\end{align}
And then, the $\eqref{}$ can be expanded by the following results
\begin{align}
\nonumber
&\frac{\kappa_{5}^{2}}{2}(\lambda+P)= \frac{2R^{\prime}\sqrt{1+B\dot{a}^{2}}}{R\sqrt{B}}\\
\label{ExpIsrael1}
&\quad \quad \quad \quad\quad+\frac{\sqrt{(1+B\dot{a}^{2})}}{2A\sqrt{B}}\big(A^{\prime}+(AB^{\prime}-A^{\prime}B)\dot{a}^{2}\big)\\
\nonumber
\\
\label{ExpIsrael2}
&\frac{\kappa_{5}^{2}}{6}(\lambda-\rho)=\frac{R^{\prime}}{R}\frac{\sqrt{1+B\dot{a}^{2}}}{\sqrt{B}}
\end{align}

\begin{figure}[ht]
	\begin{center}
		\includegraphics[scale=0.5]{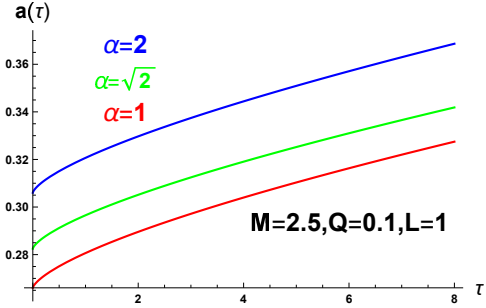}
		\caption{(color online). The variation of brane's location $a(\tau)$ in radial direction of bulk coordinates as the increase of proper time.}
		\label{Solrtau}
	\end{center}
\end{figure}

By using the Gauss-Codacci equations, the effective gravitational field on the brane is described by
\begin{align}
\nonumber
&\hspace{-1.5mm}\frac{3}{2}\bigg(\mathcal{R}_{\mu\nu}-\frac{1}{6}h_{\mu\nu}\mathcal{R}\bigg)+\frac{3}{2}E_{\mu\nu}-\frac{3}{2}\mathcal{K}\mathcal{K}_{\mu\nu}+\frac{3}{2}\mathcal{K}_{\mu\rho}\mathcal{K}_{\nu}^{\rho}+\\
\label{effEinOnbrane}
&\hspace{-1.5mm}\frac{\mathcal{K}^{2}}{4}h_{\mu\nu}-\frac{1}{4}h_{\mu\nu}\mathcal{K}_{\alpha\beta}\mathcal{K}^{\alpha\beta}=\kappa_{5}^{2}\big(\mathcal{T}_{MN}h_{\mu}^{M}h_{\nu}^{N}-\frac{\mathcal{T}}{4}h_{\mu\nu}\big)
\end{align}
where the $\mathcal{R}_{\mu\nu}$ and $\mathcal{R}$ are the Ricci tensor and Ricci scalar of the induced metric $h_{\mu\nu}$. Meanwhile the $E_{\mu\nu}$ has the following definition,
\begin{align}
\label{tidaltensor}
&E_{\mu \nu}=\mathcal{C}_{MNAB} n^M  n^A e^N _\mu e^B _\nu
\end{align}
in which the $\mathcal{C}_{MNAB}$ represents the Weyl curvature tensor in bulk spacetime, namely
\begin{align}
\nonumber
&\hspace{-3mm}\mathcal{C}_{MNAB}=R_{MNAB}-\frac{2}{3}(g_{MA}R_{BN}-g_{NA}R_{BM})+\frac{1}{6}g_{MA}g_{BN}R
\end{align} 
Substituting $\eqref{InduMetOnBrane}$, $\eqref{geneIsrael}$ into $\eqref{effEinOnbrane}$ and expanding it explicitly, we obtain two equations as follows
\begin{align}
\nonumber
&\quad\quad\frac{2R^{\prime}}{R}\ddot{a}+(\frac{A^{\prime}}{A}+\frac{B^{\prime}}{B})\frac{R^{\prime}}{R}\dot{a}^{2}+\frac{A^{\prime}}{AB}\frac{R^{\prime}}{R}-(\frac{1}{B}+1)\frac{R^{\prime2}}{R^{2}}\\
\label{EffEinOnbrane1}
&\quad=\frac{\kappa_{5}^{4}}{6}(\lambda P+\frac{\lambda^{2}}{6}-P\rho+\frac{2}{3}\lambda\rho-\frac{5}{6}\rho^{2})\\
\nonumber
\\
\label{EffEinOnbrane2}
&\quad\quad\frac{R^{\prime2}}{BR^{2}}\big(1+B\dot{a}^{2}\big)=\frac{\kappa_{5}^{4}}{18}\big(\frac{\lambda^{2}}{2}+\frac{\rho^{2}}{2}-\lambda\rho\big)
\end{align}
It should be emphasized that $R$ represents the solution $\eqref{solR}$, which is the function of $r$ in metric $\eqref{Metric}$ rather than the Ricci scalar. It is easy to observe that $\eqref{EffEinOnbrane2}$ is same with $\eqref{ExpIsrael2}$. In a word, the bulk-brane system is described by four independent equations $\eqref{DetaJunScalar},~\eqref{ExpIsrael1},~\eqref{ExpIsrael2},~\eqref{EffEinOnbrane1}$. After combining the Einstein field equation $\eqref{Eintt}$-$\eqref{Einxx}$ in bulk spacetime with the Israel junction condition $\eqref{ExpIsrael1}$-$\eqref{ExpIsrael2}$, $\eqref{EffEinOnbrane1}$ could be further simplified as
\begin{align}
\label{DeterEvort}
\ddot{a}+\frac{A^{\prime}}{2A}\dot{a}^{2}+\frac{A^{\prime}B}{2A}\dot{a}^{4}-\frac{1}{2}B^{\prime}\dot{a}^{4}=0
\end{align}
Actually, it is very difficult to find the analytical solution of $\eqref{DeterEvort}$. So, we have to consider it numerically. After substituting $\eqref{solA}$, $\eqref{solB}$ into $\eqref{DeterEvort}$ and solving this differential equation numerically at fixed parameters $M$ and $Q$ with different $\alpha$. We display the variation of brane's position $a(\tau)$ and corresponding velocity $\dot{a}(\tau)$ in Fig.\ref{Solrtau}.
Note that the initial position of brane need to satisfy $a(0)>r_h$, or the brane could not pass through the past event horizon.

\subsection{Growth rate of holographic complexity for brane-bulk system \label{ComForBBsystem}}

\begin{figure}[ht]
	\begin{center}
		\includegraphics[scale=1.15]{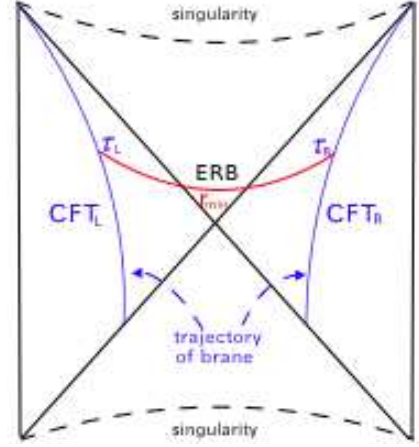}
		\caption{(color online). The trajectory of a self-graviting brane which moves in an eternal charged AdS-dilaton black holes. Note that the initial position of brane need to satisfy $a(0)>r_h$, or the brane could not pass through the past event horizon.}
		\label{BulkBranePenrose}
	\end{center}
\end{figure}

\begin{figure}[ht]
	\begin{center}
		\includegraphics[scale=0.335]{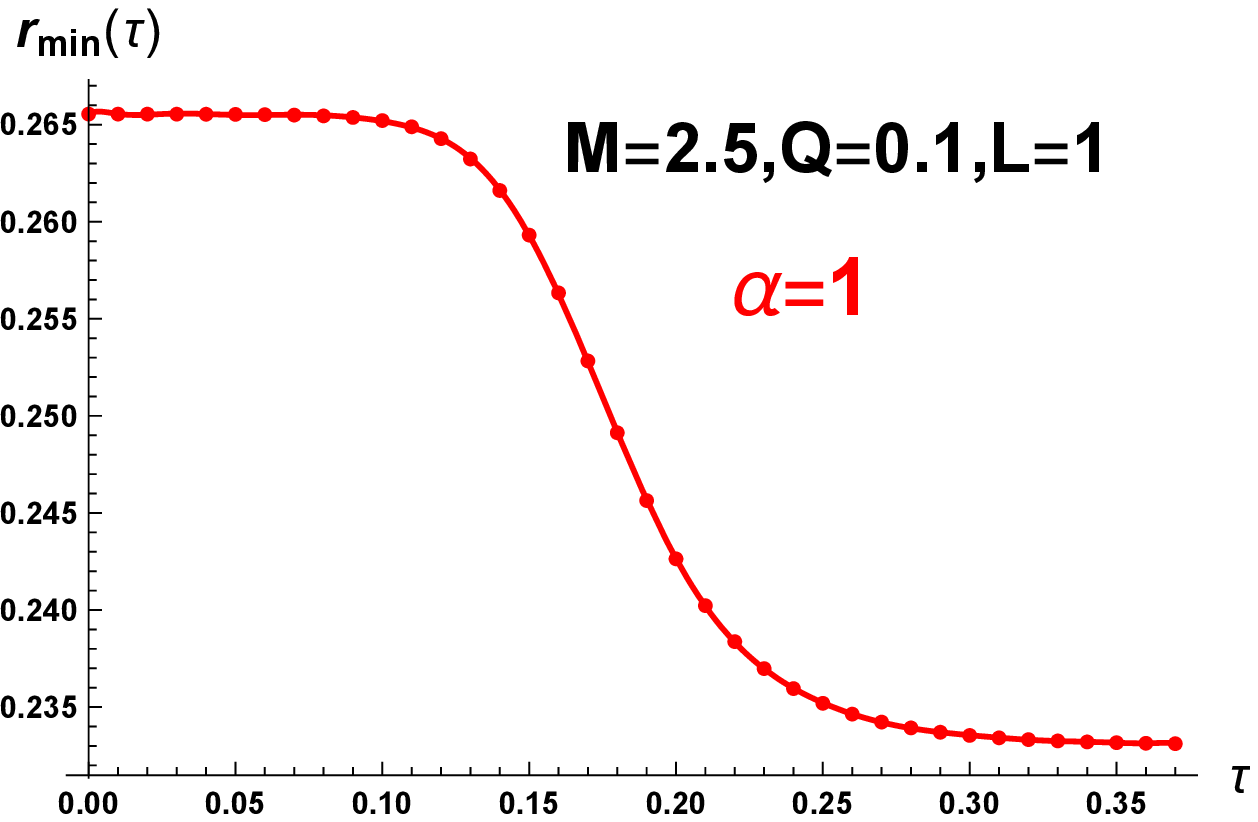}
		\includegraphics[scale=0.335]{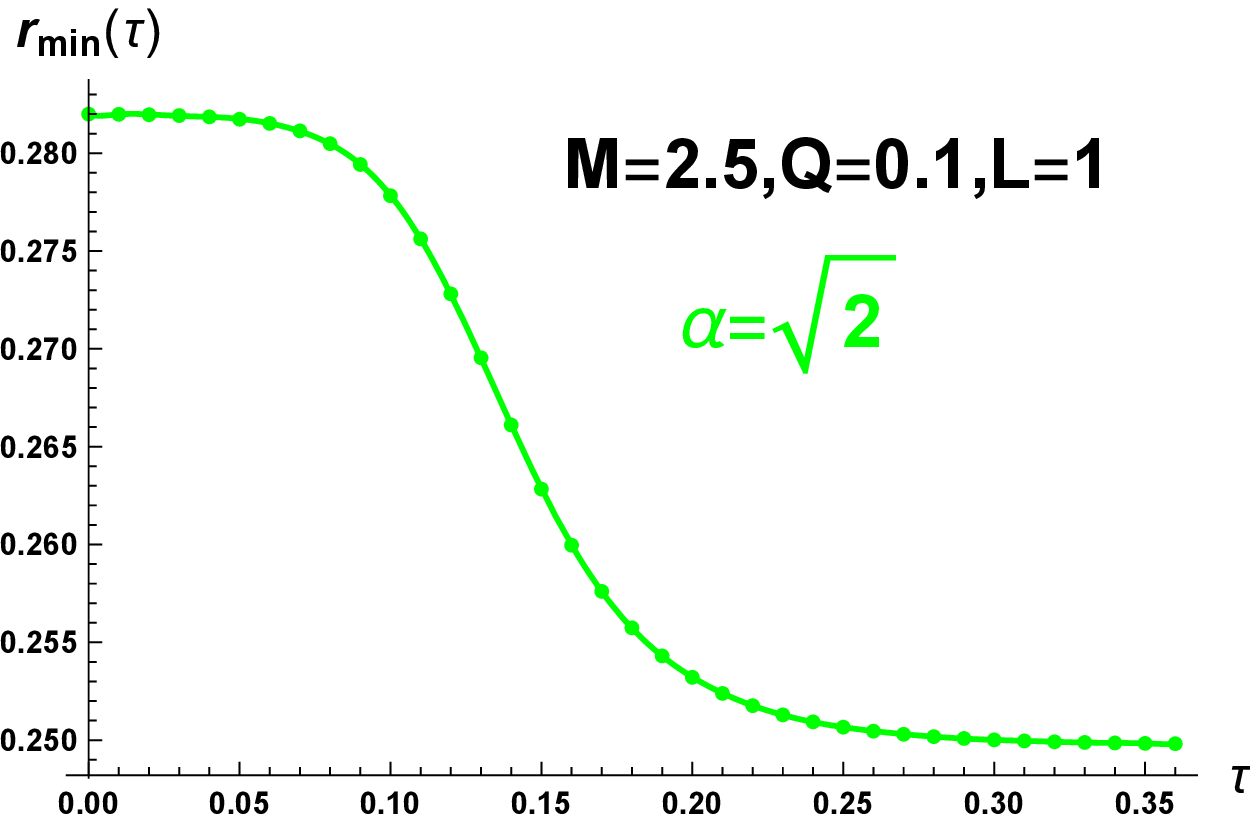}\\
		\includegraphics[scale=0.335]{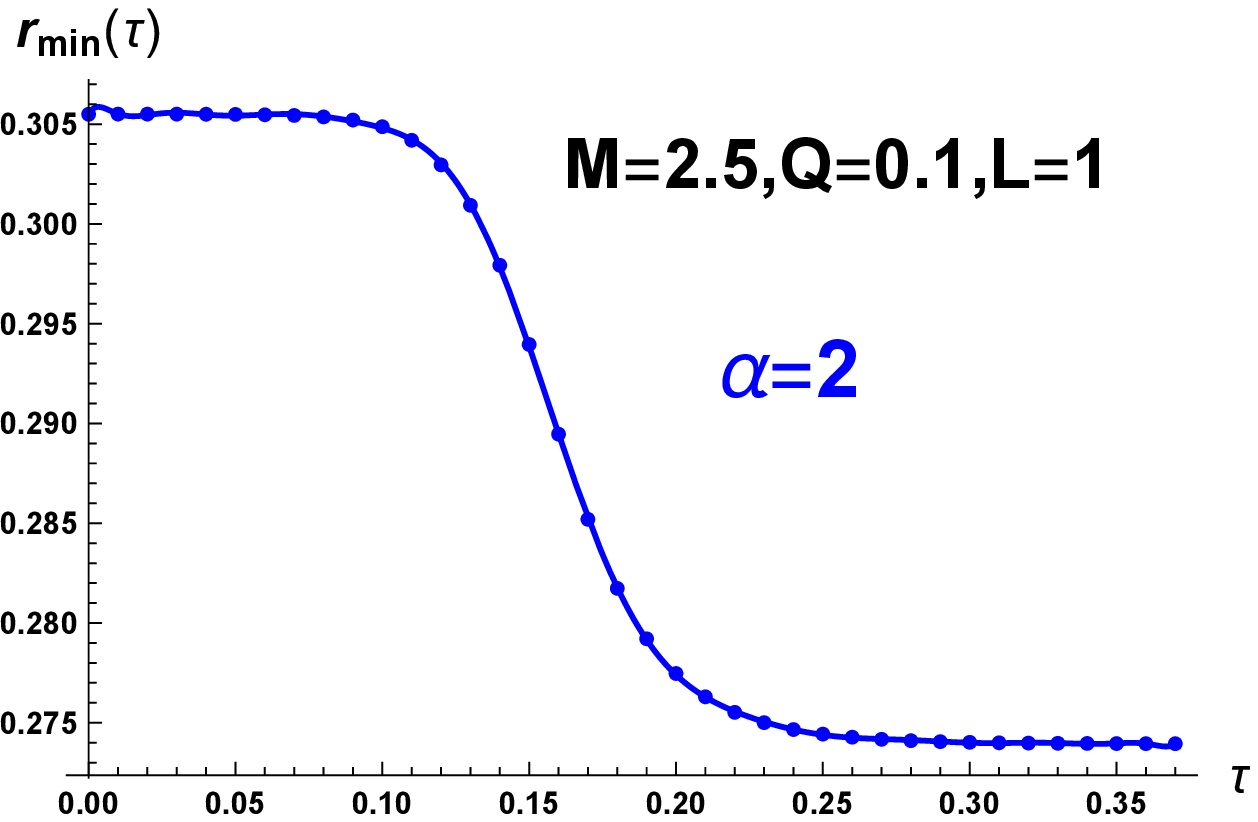}
		\includegraphics[scale=0.335]{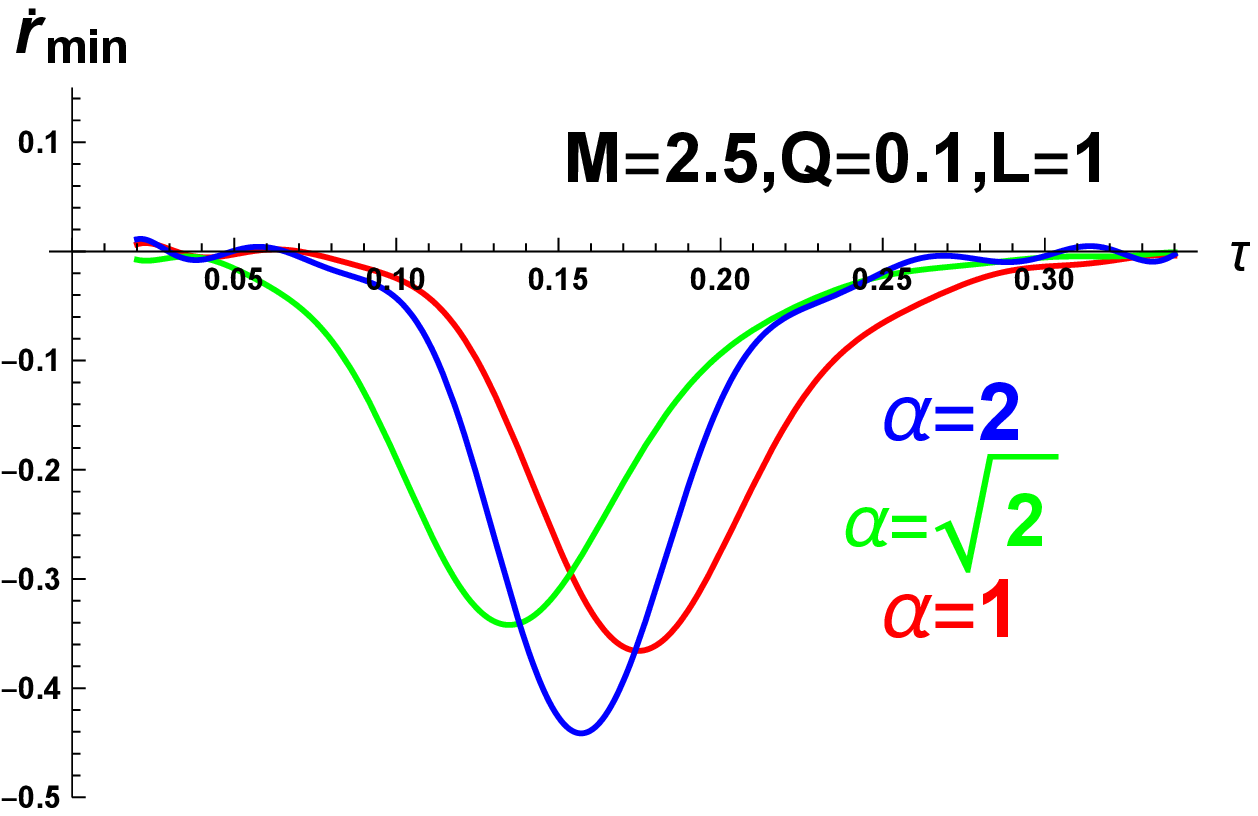}
		\caption{(color online). Find an implicit function relation between $r_{min}$ and $\tau$ through the data fitting method. The data points $\{(\tau^i,r^i_{min}),~i=1,2,\dots\}$ are collected by making $\eqref{BraneBulkTime}$ equal to the integration of $\eqref{ttauRela}$ at each $(\tau^i, r^i_{min})$. Note that the initial condition for $t(\tau)$ is $t(0)=0$, it implies $r_{min}(0)=a(0)$ (the value of $a(0)$ could be found in Fig.\ref{Solrtau}).}
		\label{rmintau}
	\end{center}
\end{figure}

As shown by Fig.\ref{BulkBranePenrose}, when the $CFT$ is located on brane, the maximal volume of wormhole $\eqref{MaxVolWorm}$ is rewritten as
\begin{align}
\label{MaxVolWormBuBra}
\mathcal{V}_{max}=2\Omega_{k,3}\int_{r_{min}}^{a(\tau)}\frac{AR^{6}}{F\sqrt{(AR^{6}+E^{2})}}dr
\end{align}
And, the time coordinates of left or right brane in bulk spacetime is 
\begin{align}
\label{BraneBulkTime}
t_{b}=\int_{r_{min}}^{a(\tau)}\big\{\frac{E}{F\sqrt{(AR^{6}+E^{2})}}\big\} dr
\end{align}
At the same time, $\eqref{BraneBulkTime}$ should also satisfy the differential equation $\eqref{ttauRela}$. Thus, an implicit function relation between $r_{min}$ and $\tau$ could be found by combining $\eqref{ttauRela}$ and $\eqref{BraneBulkTime}$. Nevertheless, it is very hard to give the analytical expression of $r_{min}(\tau)$. As shown by Fig.\ref{rmintau}, we could roughly depict the function relation between $r_{min}$ and $\tau$ by using the data fitting method.

\begin{figure}[ht]
	\begin{center}
		\includegraphics[scale=0.55]{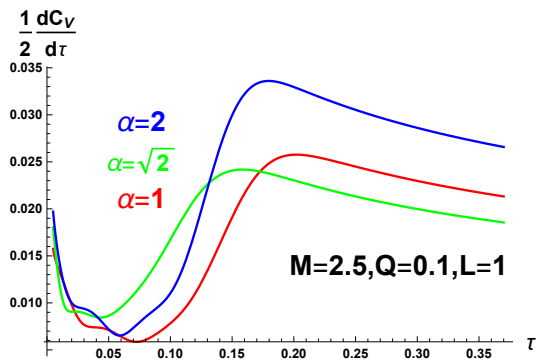}
		\caption{(color online). The full time dependence of the growth rate of the holographic complexity for the brane-bulk system in background of charged AdS-dilaton black holes. Note the $\tau$ on the horizontal axis represents the proper time on the moving brane.}
		\label{dCdtauVstau}
	\end{center}
\end{figure}

From $\eqref{MaxVolWormBuBra}$ and $\eqref{BraneBulkTime}$, one can easily prove the following equation
\begin{align}
\label{DecomMaxVol}
&\frac{\mathcal{V}_{max}}{2\Omega_{k,3}}=\int_{r_{min}}^{a(\tau)}\frac{\sqrt{(AR^{6}+E^{2})}}{F}dr-Et_{b}
\end{align}
By using the chain rule of derivatives, the growth rate of complexity with respect to $\tau$ is 
\begin{align}
\nonumber
&\hspace{-6mm}\frac{1}{2}\frac{d\mathcal{C}_{V}}{d\tau}=\frac{1}{2}\frac{\partial\mathcal{C}_{V}}{\partial t_{b}}\frac{dt_{b}}{d\tau}+\frac{1}{2}\frac{\partial\mathcal{C}_{V}}{\partial r_{min}}\frac{dr_{min}}{d\tau}+\frac{1}{2}\frac{\partial\mathcal{C}_{V}}{\partial a}\frac{da}{d\tau}\\
\nonumber
&\hspace{-3mm}\quad\quad=\frac{8\pi\Omega_{k,3}}{\kappa_{5}^{2}L}\bigg(\dot{a}(\frac{\sqrt{AR^{6}+E^{2}}}{F})_{a(\tau)}-\dot{r}_{min}(\frac{\sqrt{AR^{6}+E^{2}}}{F})_{r_{min}}\\
\nonumber
&\hspace{-3mm}\quad\quad\quad~+\dot{r}_{min}\int_{r_{min}}^{a(\tau)}\frac{\partial}{\partial r_{min}}(\frac{\sqrt{AR^{6}+E^{2}}}{F})dr\bigg)\\
\label{ComGrowRate0}
&\hspace{-3mm}\quad\quad\quad~-\frac{8\pi\Omega_{k,3}}{\kappa_{5}^{2}L}\bigg(E\frac{dt_{b}}{d\tau} +\frac{\partial E}{\partial r_{min}}\dot{r}_{min}t_{b}\bigg)
\end{align}
in which the factor $\frac{1}{2}$ is due to the fact that the time dependence of the complexity are only relevant to the total time $\tau_L+\tau_R=2\tau$. By utilizing the following results
\begin{align}
\nonumber
&A(r_{min})R^{6}(r_{min})+E^{2}=0\\
\nonumber
\\
\nonumber
&\int_{r_{min}}^{a(\tau)}\frac{\partial}{\partial r_{min}}(\frac{\sqrt{AR^{6}+E^{2}}}{F})dr=\frac{\partial E}{\partial r_{min}}t_{b}
\end{align}
The $\eqref{ComGrowRate0}$ can be simplified as
\begin{align}
\label{ComGrowRate1}
\frac{1}{2}\frac{d\mathcal{C}_{V}}{d\tau}=\frac{8\pi\Omega_{k,3}}{\kappa_{5}^{2}L}\bigg(\dot{a}(\frac{\sqrt{AR^{6}+E^{2}}}{F})_{a(\tau)}-E\frac{dt_{b}}{d\tau}\bigg)
\end{align}
By putting the numerical solutions $r_{min}(\tau)$ and $a(\tau)$ into $\eqref{ComGrowRate1}$, the growth rate of complexity for CFT located on brane is plotted in Fig.\ref{dCdtauVstau}
It is straightforward to see that the growth rate decreases monotonously on late time, after arriving at a maximum value. This variation trend is similiar to the complexity evolution for the closed brane universe via the CV conjecture given by \cite{An:2019opz}, but there is no negative value in our result. This difference is easy to understand from the expression $\eqref{ComGrowRate1}$. As studied by the \cite{An:2019opz}, the evolution mode of the closed brane universe is in a contracting phase on late time, namely $\dot{a}<0$. Thus, the $\frac{d\mathcal{C}_{V}}{d\tau}<0$ is obtaind naturally if $\dot{a}$ is negative. But in our case, as shown by Fig.\ref{Solrtau}, the growth rate is positive due to $\dot{a}>0$. By analyzing the results in \cite{An:2019opz} and our work, we find that the growth rate of holographic complexity for brane-bulk system on late time is dominated by the velocity of the brane. Meanwhile, we guess this conclusion is model-independent. But the impact factors to the growth rate of complexity on early time is still unclear, we need more case studies to uncover it.

\section{Conclusion and Discussion \label{ConAndDis}}

In this paper, we have investigated the growth rate of holographic complexity for charged AdS-dilaton black holes in a Einstein-Maxwell-Dilaton gravity by using the so called complexity-volume (CV) conjecture which associates the quantum complexity of the CFT on boundary to the maximum volume of the ERB in bulk, as illustrated by Fig.\ref{AdSDilaPenrose}. At first, we consider the case of static black hole bulk with a fixed boundary which is located on the infinity. In Fig.\ref{PicGrowCom}, we plot the full time dependence of the growth rate of holographic complexity with different dilaton coupling constant $\alpha$. It is easy to observe that there exists a bound for growth rate on late times, while this bound will become larger as we increase the value of $\alpha$. For ensuring that the bound is a finite value when $\alpha\to\infty$, we calculate it analytically in large $\alpha$ limit. And, the result $\eqref{FinaBound}$ means that the complexity growth rate, on the late time, can not exceed the bound $\frac{8\pi M}{\sqrt{3}}$, whatever how large the dilaton coupling $\alpha$ is. Our results are complementary to the \cite{An:2018xhv}, in which the CA proposal for calculation of holographic complexity grwoth in background of a charged AdS-dilaton black holes has been considered.

On the next step, we replace the AdS boundary of black holes by a moving self-graviting RS brane, while the growth rate of holographic complexity for this brane-bulk system is analyzed. Specifically, the movement of brane is determined by the Israel junction condition and the effective Einstein field equation on the brane together. We derive the effective Einstein field equation on the brane by using the so called projection method provided by \cite{Shiromizu:1999wj, Maeda:2003vq}. Meanwhile the analysis about the brane's motion mainly refer to the results given by \cite{Li:2019jal}. Due to the self-gravitating effects of brane, in Fig.\ref{Solrtau}, we observe the evolution mode of the brane is not sensitive to the value of dilaton coupling constant. On the other hand, as shown by trajectory of the brane in Penrose diagram Fig.\ref{BulkBranePenrose}, the initial position of brane need to fulfill $a(0)>r_h$, or the brane could not cross the past event horizon. Note that if we turn off the self-gravitating effects, the evolution mode of the brane will change dramatically as varying the value of $\alpha$ \cite{Chamblin:1999ya, Xu:2019abl}, meanwhile the brane will run from the initial position near the singularity of white hole \cite{Savonije:2001nd, Padilla:2002tg}. Basing on Fig.\ref{BulkBranePenrose}, the growth rate of holographic complexity for this brane-bulk system via the CV conjecture is calculated in Sec.\ref{ComForBBsystem}. From $\eqref{ComGrowRate1}$, the time evolution behavior of the growth rate of holographic complexity for this brane-bulk system in different $\alpha$ is displayed by Fig.\ref{dCdtauVstau}. It is straightforward to see that the growth rate decreases monotonously on late time, after arriving at a maximum value. This variation trend is similiar to the complexity evolution for the closed brane universe via the CV conjecture given by \cite{An:2019opz}, but there is no negative value in our result.

As discussions, we suggest the following extensions. The first case is, investigate the holographic complexity growth via the CA conjecture for this brane-bulk system in Einstein-Maxwell-Dilaton gravity. And it is important to check that if the full time dependence of the growth rate in CA conjecture is consistent with the one obtained in CV conjecture. Besides, \cite{Bhattacharyya:2020rpy} considers the time dependence of the quantum computational complexity of scalar curvature perturbations on an early-time period of de Sitter expansion followed by a radiation-dominated era. And it is interesting to generalize the scalar curvature perturbations into the curvaton models\cite{Enqvist:2001zp, Lyth:2001nq, Moroi:2001ct, Liu:2019xhn, Liu:2020zzv}. Finally, it is also worthwhile to explore the holographic complexity growth for AdS-dilaton black holes with a probe string, like the work \cite{Nagasaki:2017kqe, Nagasaki:2019icm}.

\section{Acknowledgements}

This work is supported by The Center for Research and Development in Mathematics and Applications (CIDMA) through the Portuguese 
Foundation for Science and Technology (FCT - Fundação para a Ciência e a Tecnologia), references UIDB/04106/2020 and UIDP/04106/2020. Besides, AC is also supported by NSFC grant no.11875082.


\begin{thebibliography}{99}

\bibitem{Maldacena:1997re} 
J.~M.~Maldacena,
``The Large N limit of superconformal field theories and supergravity,''
{\it Int. J. Theor. Phys.} {\bf 38}, 1113 (1999)
{\it Adv. Theor. Math. Phys.} {\bf 2}, 231 (1998),
\href{https://arxiv.org/abs/hep-th/9711200}{arXiv: hep-th/9711200}

\bibitem{Witten:1998qj}
E.~Witten,
``Anti-de Sitter space and holography,''
{\it Adv. Theor. Math. Phys.} {\bf 2}, 253 (1998),
\href{https://arxiv.org/abs/hep-th/9802150}{arXiv: hep-th/9802150}

\bibitem{Gubser:1998bc}
S.~Gubser, I.~R.~Klebanov and A.~M.~Polyakov,
``Gauge theory correlators from noncritical string theory,''
Phys. Lett. B \textbf{428} (1998), 105-114
doi:10.1016/S0370-2693(98)00377-3
[arXiv:hep-th/9802109 [hep-th]].

\bibitem{Hartnoll:2008vx}
S.~A.~Hartnoll, C.~P.~Herzog and G.~T.~Horowitz,
``Building a Holographic Superconductor,''
Phys. Rev. Lett. \textbf{101} (2008), 031601
doi:10.1103/PhysRevLett.101.031601
[arXiv:0803.3295 [hep-th]].

\bibitem{Gubser:2008yx}
S.~S.~Gubser, A.~Nellore, S.~S.~Pufu and F.~D.~Rocha,
``Thermodynamics and bulk viscosity of approximate black hole duals to finite temperature quantum chromodynamics,''
Phys. Rev. Lett. \textbf{101} (2008), 131601
doi:10.1103/PhysRevLett.101.131601
[arXiv:0804.1950 [hep-th]].

\bibitem{Hartnoll:2009sz}
S.~A.~Hartnoll,
``Lectures on holographic methods for condensed matter physics,''
Class. Quant. Grav. \textbf{26} (2009), 224002
doi:10.1088/0264-9381/26/22/224002
[arXiv:0903.3246 [hep-th]].

\bibitem{Ryu:2006bv}
S.~Ryu and T.~Takayanagi,
``Holographic derivation of entanglement entropy from AdS/CFT,''
Phys. Rev. Lett. \textbf{96} (2006), 181602
doi:10.1103/PhysRevLett.96.181602
[arXiv:hep-th/0603001 [hep-th]].

\bibitem{Casini:2011kv}
H.~Casini, M.~Huerta and R.~C.~Myers,
``Towards a derivation of holographic entanglement entropy,''
JHEP \textbf{05} (2011), 036
doi:10.1007/JHEP05(2011)036
[arXiv:1102.0440 [hep-th]].

\bibitem{Maldacena:2001kr}
J.~M.~Maldacena,
``Eternal black holes in anti-de Sitter,''
JHEP \textbf{04} (2003), 021
doi:10.1088/1126-6708/2003/04/021
[arXiv:hep-th/0106112 [hep-th]].

\bibitem{Hartman:2013qma}
T.~Hartman and J.~Maldacena,
``Time Evolution of Entanglement Entropy from Black Hole Interiors,''
JHEP \textbf{05} (2013), 014
doi:10.1007/JHEP05(2013)014
[arXiv:1303.1080 [hep-th]].

\bibitem{Stanford:2014jda}
D.~Stanford and L.~Susskind,
``Complexity and Shock Wave Geometries,''
Phys. Rev. D \textbf{90}, no.12, 126007 (2014)
doi:10.1103/PhysRevD.90.126007
[arXiv:1406.2678 [hep-th]].

\bibitem{Susskind:2014rva}
L.~Susskind,
``Computational Complexity and Black Hole Horizons,''
Fortsch. Phys. \textbf{64} (2016), 24-43
doi:10.1002/prop.201500092
[arXiv:1403.5695 [hep-th]].

\bibitem{Brown:2015bva}
A.~R.~Brown, D.~A.~Roberts, L.~Susskind, B.~Swingle and Y.~Zhao,
``Holographic Complexity Equals Bulk Action?,''
Phys. Rev. Lett. \textbf{116} (2016) no.19, 191301
doi:10.1103/PhysRevLett.116.191301
[arXiv:1509.07876 [hep-th]].

\bibitem{Brown:2015lvg}
A.~R.~Brown, D.~A.~Roberts, L.~Susskind, B.~Swingle and Y.~Zhao,
``Complexity, action, and black holes,''
Phys. Rev. D \textbf{93} (2016) no.8, 086006
doi:10.1103/PhysRevD.93.086006
[arXiv:1512.04993 [hep-th]].

\bibitem{Cai:2016xho}
R.~G.~Cai, S.~M.~Ruan, S.~J.~Wang, R.~Q.~Yang and R.~H.~Peng,
``Action growth for AdS black holes,''
JHEP \textbf{09} (2016), 161
doi:10.1007/JHEP09(2016)161
[arXiv:1606.08307 [gr-qc]].

\bibitem{Cai:2017sjv}
R.~G.~Cai, M.~Sasaki and S.~J.~Wang,
``Action growth of charged black holes with a single horizon,''
Phys. Rev. D \textbf{95} (2017) no.12, 124002
doi:10.1103/PhysRevD.95.124002
[arXiv:1702.06766 [gr-qc]].

\bibitem{Alishahiha:2017hwg}
M.~Alishahiha, A.~Faraji Astaneh, A.~Naseh and M.~H.~Vahidinia,
``On complexity for F(R) and critical gravity,''
JHEP \textbf{05} (2017), 009
doi:10.1007/JHEP05(2017)009
[arXiv:1702.06796 [hep-th]].

\bibitem{Cano:2018aqi}
P.~A.~Cano, R.~A.~Hennigar and H.~Marrochio,
``Complexity Growth Rate in Lovelock Gravity,''
Phys. Rev. Lett. \textbf{121} (2018) no.12, 121602
doi:10.1103/PhysRevLett.121.121602
[arXiv:1803.02795 [hep-th]].

\bibitem{Jiang:2018pfk}
J.~Jiang,
``Action growth rate for a higher curvature gravitational theory,''
Phys. Rev. D \textbf{98} (2018) no.8, 086018
doi:10.1103/PhysRevD.98.086018
[arXiv:1810.00758 [hep-th]].

\bibitem{An:2018dbz}
Y.~S.~An, R.~G.~Cai and Y.~Peng,
``Time Dependence of Holographic Complexity in Gauss-Bonnet Gravity,''
Phys. Rev. D \textbf{98} (2018) no.10, 106013
doi:10.1103/PhysRevD.98.106013
[arXiv:1805.07775 [hep-th]].

\bibitem{Miao:2017quj}
Y.~G.~Miao and L.~Zhao,
``Complexity-action duality of the shock wave geometry in a massive gravity theory,''
Phys. Rev. D \textbf{97} (2018) no.2, 024035
doi:10.1103/PhysRevD.97.024035
[arXiv:1708.01779 [hep-th]].

\bibitem{Pan:2016ecg}
W.~J.~Pan and Y.~C.~Huang,
``Holographic complexity and action growth in massive gravities,''
Phys. Rev. D \textbf{95} (2017) no.12, 126013
doi:10.1103/PhysRevD.95.126013
[arXiv:1612.03627 [hep-th]].

\bibitem{Brown:2017jil}
A.~R.~Brown and L.~Susskind,
``Second law of quantum complexity,''
Phys. Rev. D \textbf{97} (2018) no.8, 086015
doi:10.1103/PhysRevD.97.086015
[arXiv:1701.01107 [hep-th]].

\bibitem{Carmi:2017jqz}
D.~Carmi, S.~Chapman, H.~Marrochio, R.~C.~Myers and S.~Sugishita,
``On the Time Dependence of Holographic Complexity,''
JHEP \textbf{11}, 188 (2017)
doi:10.1007/JHEP11(2017)188
[arXiv:1709.10184 [hep-th]].

\bibitem{Mahapatra:2018gig}
S.~Mahapatra and P.~Roy,
``On the time dependence of holographic complexity in a dynamical Einstein-dilaton model,''
JHEP \textbf{11} (2018), 138
doi:10.1007/JHEP11(2018)138
[arXiv:1808.09917 [hep-th]].

\bibitem{Carmi:2016wjl}
D.~Carmi, R.~C.~Myers and P.~Rath,
``Comments on Holographic Complexity,''
JHEP \textbf{03} (2017), 118
doi:10.1007/JHEP03(2017)118
[arXiv:1612.00433 [hep-th]].

\bibitem{Reynolds:2016rvl}
A.~Reynolds and S.~F.~Ross,
``Divergences in Holographic Complexity,''
Class. Quant. Grav. \textbf{34} (2017) no.10, 105004
doi:10.1088/1361-6382/aa6925
[arXiv:1612.05439 [hep-th]].

\bibitem{Chapman:2018dem}
S.~Chapman, H.~Marrochio and R.~C.~Myers,
``Holographic complexity in Vaidya spacetimes. Part I,''
JHEP \textbf{06} (2018), 046
doi:10.1007/JHEP06(2018)046
[arXiv:1804.07410 [hep-th]].

\bibitem{Chapman:2018lsv}
S.~Chapman, H.~Marrochio and R.~C.~Myers,
``Holographic complexity in Vaidya spacetimes. Part II,''
JHEP \textbf{06} (2018), 114
doi:10.1007/JHEP06(2018)114
[arXiv:1805.07262 [hep-th]].

\bibitem{Jiang:2018tlu}
J.~Jiang,
``Holographic complexity in charged Vaidya black hole,''
Eur. Phys. J. C \textbf{79} (2019) no.2, 130
doi:10.1140/epjc/s10052-019-6639-1
[arXiv:1811.07347 [hep-th]].

\bibitem{Reynolds:2017lwq}
A.~Reynolds and S.~F.~Ross,
``Complexity in de Sitter Space,''
Class. Quant. Grav. \textbf{34} (2017) no.17, 175013
doi:10.1088/1361-6382/aa8122
[arXiv:1706.03788 [hep-th]].

\bibitem{Caginalp:2019fyt}
R.~J.~Caginalp,
``Holographic Complexity in FRW Spacetimes,''
Phys. Rev. D \textbf{101} (2020) no.6, 066027
doi:10.1103/PhysRevD.101.066027
[arXiv:1906.02227 [hep-th]].

\bibitem{An:2019opz}
Y.~S.~An, R.~G.~Cai, L.~Li and Y.~Peng,
``Holographic complexity growth in a FLRW universe,''
\href{https://arxiv.org/abs/1909.12172}{arXiv:1909.12172 [hep-th]}

\bibitem{Pan:2020kdl}
W.~J.~Pan, Y.~l.~Li, M.~Song, W.~b.~Xie and S.~Zhang,
``Holographic Complexity Growth Rate in a dual FLRW Universe,''
[arXiv:2003.11415 [hep-th]].

\bibitem{Gibbons:1987ps}
G.~W.~Gibbons and K.~i.~Maeda,
``Black Holes and Membranes in Higher Dimensional Theories with Dilaton Fields,''
Nucl.\ Phys.\ B {\bf 298} (1988) 741.
doi:10.1016/0550-3213(88)90006-5

\bibitem{Gao:2004tv}
C.~J.~Gao and S.~N.~Zhang,
``Higher dimensional dilaton black holes with cosmological constant,''
Phys.\ Lett.\ B {\bf 605} (2005) 185
doi:10.1016/j.physletb.2004.11.030
\href{https://arxiv.org/abs/hep-th/0411105}{hep-th/0411105}

\bibitem{Gao:2005xv}
C.~J.~Gao and S.~N.~Zhang,
``Topological black holes in dilaton gravity theory,''
Phys.\ Lett.\ B {\bf 612} (2005) 127.
doi:10.1016/j.physletb.2005.03.026

\bibitem{Randall:1999ee}
L.~Randall and R.~Sundrum,
``A Large mass hierarchy from a small extra dimension,''
{\it Phys.\ Rev.\ Lett.\ }  {\bf 83} (1999) 3370
\href{https://arxiv.org/abs/hep-ph/9905221}{arXiv: hep-ph/9905221}

\bibitem{Randall:1999vf}
L.~Randall and R.~Sundrum,
``An Alternative to compactification,''
{\it Phys.\ Rev.\ Lett.\ } {\bf 83} (1999) 4690
\href{https://arxiv.org/abs/hep-th/9906064}{arXiv: hep-th/9906064}

\bibitem{Chamblin:1999ya}
H.~A.~Chamblin and H.~S.~Reall,
``Dynamic dilatonic domain walls,''
{\it Nucl.\ Phys.\ B} {\bf 562}, 133 (1999)
\href{https://arxiv.org/abs/hep-th/9903225}{arXiv:9903225 [hep-th]}

\bibitem{Kraus:1999it}
P.~Kraus,
``Dynamics of anti-de Sitter domain walls,''
JHEP {\bf 9912} (1999) 011
doi:10.1088/1126-6708/1999/12/011
\href{https://arxiv.org/abs/hep-th/9910149}{arXiv:hep-th/9910149}

\bibitem{Sheykhi:2009pf}
A.~Sheykhi, M.~H.~Dehghani and S.~H.~Hendi,
``Thermodynamic instability of charged dilaton black holes in AdS spaces,''
Phys. Rev. D \textbf{81} (2010), 084040
doi:10.1103/PhysRevD.81.084040
[arXiv:0912.4199 [hep-th]].

\bibitem{Li:2017kkj}
A.~c.~Li, H.~q.~Shi and D.~f.~Zeng,
``Phase structure and quasinormal modes of a charged AdS dilaton black hole,''
Phys.\ Rev.\ D {\bf 97} (2018) no.2,  026014
doi:10.1103/PhysRevD.97.026014
\href{https://arxiv.org/abs/1711.04613}{arXiv:1711.04613 [hep-th]}

\bibitem{Xu:2019abl} 
W.~L.~Xu and Y.~C.~Huang,
``Dynamics of domain wall in charged AdS dilaton black hole space–time,''
Int.\ J.\ Mod.\ Phys.\ A {\bf 34} (2019) no.23,  1950132
doi:10.1142/S0217751X1950132X
\href{https://arxiv.org/abs/1905.03688}{arXiv:1905.03688 [gr-qc]}

\bibitem{Li:2019jal}
A.~c.~Li,
``Brane Universe and Holography in Spacetime of Charged AdS Dilaton Black Hole,''
Phys. Rev. D \textbf{101} (2020) no.8, 086019
doi:10.1103/PhysRevD.101.086019
[arXiv:1912.11319 [hep-th]].

\bibitem{Shiromizu:1999wj}
T.~Shiromizu, K.~i.~Maeda and M.~Sasaki,
``The Einstein equation on the 3-brane world,''
{\it Phys.\ Rev.\ D} {\bf 62} (2000) 024012
\href{https://arxiv.org/abs/gr-qc/9910076}{arXiv: gr-qc/9910076}

\bibitem{Maeda:2003vq}
K.~i.~Maeda and T.~Torii,
``Covariant gravitational equations on brane world with Gauss-Bonnet term,''
Phys. Rev. D \textbf{69} (2004), 024002
doi:10.1103/PhysRevD.69.024002
[arXiv:hep-th/0309152 [hep-th]].

\bibitem{Balasubramanian:1999re}
V.~Balasubramanian and P.~Kraus,
``A Stress tensor for Anti-de Sitter gravity,''
{\it Commun.\ Math.\ Phys.\  }{\bf 208} (1999) 413
doi:10.1007/s002200050764
\href{https://arxiv.org/abs/hep-th/9902121}{arXiv:hep-th/9902121}

\bibitem{Cai:1999xg}
R.~G.~Cai and N.~Ohta,
``Surface counterterms and boundary stress energy tensors for asymptotically nonAnti-de Sitter spaces,''
Phys.\ Rev.\ D {\bf 62} (2000) 024006
doi:10.1103/PhysRevD.62.024006
\href{https://arxiv.org/abs/hep-th/9912013}{arXiv:hep-th/9912013}

\bibitem{Savonije:2001nd}
I.~Savonije and E.~P.~Verlinde,
``CFT and entropy on the brane,''
Phys.\ Lett.\ B {\bf 507} (2001) 305
doi:10.1016/S0370-2693(01)00467-1
\href{https://arxiv.org/abs/hep-th/0102042}{arXiv: hep-th/0102042}

\bibitem{Padilla:2002tg}
A.~Padilla,
``Brane world cosmology and holography,''
[arXiv:hep-th/0210217 [hep-th]].

\bibitem{An:2018xhv}
Y.~S.~An and R.~H.~Peng,
``Effect of the dilaton on holographic complexity growth,''
Phys. Rev. D \textbf{97} (2018) no.6, 066022
doi:10.1103/PhysRevD.97.066022
[arXiv:1801.03638 [hep-th]].

\bibitem{Bhattacharyya:2020rpy}
A.~Bhattacharyya, S.~Das, S.~Shajidul Haque and B.~Underwood,
``Cosmological Complexity,''
Phys. Rev. D \textbf{101} (2020) no.10, 106020
doi:10.1103/PhysRevD.101.106020
[arXiv:2001.08664 [hep-th]].

\bibitem{Enqvist:2001zp}
K.~Enqvist and M.~S.~Sloth,
``Adiabatic CMB perturbations in pre - big bang string cosmology,''
Nucl. Phys. B \textbf{626} (2002), 395-409
doi:10.1016/S0550-3213(02)00043-3
[arXiv:hep-ph/0109214 [hep-ph]].

\bibitem{Lyth:2001nq}
D.~H.~Lyth and D.~Wands,
``Generating the curvature perturbation without an inflaton,''
Phys. Lett. B \textbf{524} (2002), 5-14
doi:10.1016/S0370-2693(01)01366-1
[arXiv:hep-ph/0110002 [hep-ph]].

\bibitem{Moroi:2001ct}
T.~Moroi and T.~Takahashi,
``Effects of cosmological moduli fields on cosmic microwave background,''
Phys. Lett. B \textbf{522} (2001), 215-221
doi:10.1016/S0370-2693(01)01295-3
[arXiv:hep-ph/0110096 [hep-ph]].

\bibitem{Liu:2019xhn}
L.~H.~Liu and W.~L.~Xu,
``The running curvaton,''
[arXiv:1911.10542 [astro-ph.CO]].

\bibitem{Liu:2020zzv}
L.~H.~Liu and T.~Prokopec,
``Non-minimally coupled curvaton,''
[arXiv:2005.11069 [astro-ph.CO]].

\bibitem{Nagasaki:2017kqe}
K.~Nagasaki,
``Complexity of AdS$_5$ black holes with a rotating string,''
Phys. Rev. D \textbf{96} (2017) no.12, 126018
doi:10.1103/PhysRevD.96.126018
[arXiv:1707.08376 [hep-th]].

\bibitem{Nagasaki:2019icm}
K.~Nagasaki,
``Complexity Growth for Topological Black Holes with a Probe String,''
[arXiv:1912.03567 [hep-th]].


\end{thebibliography}
\end{document}